\documentstyle[epsf]{aipproc}

\def\grs{GRS~1915+105}
\def\gro{GRO~J1655-40}
\def\rxte{{\it RXTE}}
\def\msol{M_\odot}
\def\psec{s$^{-1}$}

\begin{document}

\title{Disk and Corona Instabilities in GRS~1915+105}

\author{Jean Swank$^1$, Xingming Chen$^2$, Craig Markwardt$^1$, 
Ronald Taam$^3$}
\address{$^1$ NASA/GSFC Code 662, Greenbelt, MD 20771\\
         $^2$ UCO/Lick Observatory, UCSC, Santa Cruz, CA 95064\\
	 $^3$ Department of Physics and Astronomy, Northwestern University,\\
	      Evanston, IL 60208}

%\lefthead{LEFT head}
%\righthead{RIGHT head}
\maketitle

\begin{abstract}

We present time-resolved GRS~1915+105 energy and power spectra
observed by \rxte, during an episode where the X-ray intensity makes
an extreme dip. If the spectra are modeled in terms of disk and power
law components, both have large variations. When the inner disk is
disrupted, the power law dominates, exhibiting quasi-periodic
oscillations with varying frequency until the inner disk returns.

\end{abstract}

\section*{Introduction}

The source \grs\ was first discovered in 1992 as an X-ray transient by
GRANAT/Watch \cite{cbl92} Subsequent radio observations showed that it
produces superluminal jet-like outflows \cite{mr94}.  While the mass
of \grs\ has not yet been determined, its behavior is similar to that
of another recently discovered jet-producing transient, \gro, whose
mass has been quite accurately determined to be $7 \msol$ \cite{ob97},
well above the mass limit for neutron stars.  \grs\ can also be quite
luminous ($L \ge 10^{39}$~erg~\psec), above the Eddington limit for
neutron stars.  Hence, both \grs\ and \gro\ are thought to be
jet-producing black hole systems with accretion disks, and have been
called ``galactic microquasars'' in reference to their AGN
counterparts.

\grs\ is a very bright X-ray source, and there are copious X-ray
observations by \rxte\ showing that the source has a rich variety of
states \cite{mrg97,cst97}, and a weak 67~Hz QPO which is presumably
associated with the inner edge of the accretion disk.  We present here
time-resolved spectroscopy of \grs\ as it enters and recovers from a
disk disruption episode.

% The precise nature of the states,
% and the transition mechanism are not well understood, but there are
% some aspects of the spectral and temporal behavior which are beginning
% to clarify.  To be specific, there are certain X-ray ``dips'' where
% the inner portion of the putative accretion disk seems to vanish.

\begin{figure}[tbp]
\centerline{\epsfxsize=\textwidth\epsffile{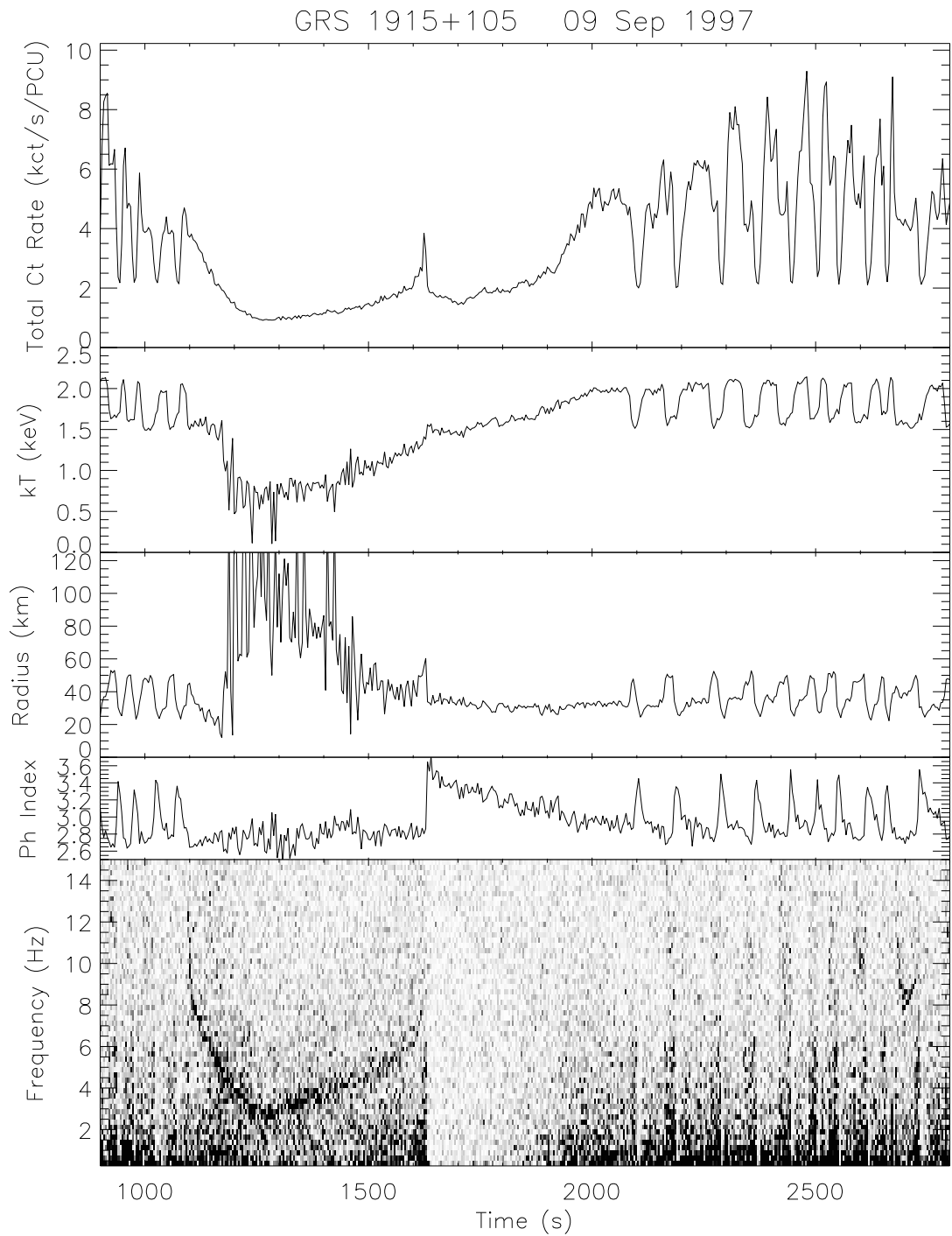}}
\caption{\rxte\ observation of \grs\ showing from top to bottom, the
X-ray light curve, inner disk temperature, inner disk radius, photon
index, and power density spectrum. \label{xte1915fig1}}
\end{figure}

%\section*{Observations and Results}

We have analyzed \rxte\ PCA data taken on 09~Sep~1997.  The source was
in a state characterized by rapid oscillations ($\sim 50$~s), followed
by extreme dips in the X-ray intensity lasting several hundred
seconds.  In order to investigate the dynamics, data in various PCA
modes were combined to construct energy spectra and power spectra at
4~s intervals.

\section*{Spectral Variations}

The energy spectra were fitted by model emission consisting of a
multicolor ``disk'' black body \cite{m84} representing emission from
an optically thick accretion disk, and a power law, which may be
associated with Compton scattering from energetic electrons. The
fitting of the disk model provides information about the temperature
and radius at the inner edge of the accretion disk, which are plotted
in the second and third panels of Figure~\ref{xte1915fig1}.  It should
be noted that no corrections for electron scattering or general
relativistic effects have been applied to the calculation of the disk
radius \cite{zcc97}, so the plotted values should be viewed as an
underestimate by a factor of about 3.  The power law photon index is
displayed in the fourth panel.  As can be seen from the plots, the
dynamics of the system is quite complex.

One remarkable feature is the extreme dip that occurs from about
1100~s to 1600~s where the X-ray flux, bracketed by rapid X-ray
intensity oscillations, drops by a factor of about eight.  Such dips
have been noted previously \cite{gmr96,b97}.  The temperature of the
disk drops and the radius of the optically thick inner edge increases
by a factor of about 5 (the signal is noisy because of the low
temperature).  A natural conclusion \cite{b97}, is that the inner
portion of the disk is disrupted during the dips: the soft X-rays that
remain are interpreted as emission from a residual disk with inner
radius of several hundred kilometers.

% The data are not particularly good during the
% most extreme part of the dip, where the black body parameters are
% poorly constrained due to the \rxte\ response, but it should be clear
% that the inner part of the accretion disk is disrupted on timescales
% of a few tens of seconds.  The minimum inner disk radius is consistent
% with the radius of marginally stable orbits found from the 67~Hz QPO
% (Morgan et al 1997).

We note here that the dips appear to be strongly associated with
relatively harder power law emission; the onset of the dip is
foreshadowed by a decrease of the power law photon index (i.e.,
hardening) near time 1100~s, and the recovery of the disk occurs only
after the power law index has also recovered, sharply at 1600~s in
Figure~\ref{xte1915fig1}. The power law flux itself (not shown) is
also much higher during the dip.  As the power law component is
commonly thought to represent inverse Compton emission from a corona
of hot electrons above the disk and/or closer to the black hole, it
seems clear that the corona is varying rapidly and is quite closely
coupled to the accretion disk.

The small spike that appears in the X-ray light curve near 1600~s
signals a dramatic change in the power law component of \grs, while it
is only weakly apparent in the black body parameters.  It precedes a
period where the accretion disk gradually recovers.  In this
observation, the spike is very distinct; in some early observations it
is harder to distinguish, as the rapid oscillations begin very soon
after the spike.  The spike also appears to be associated with
radio-emitting outflows.  Simultaneous observations by Mirabel et
al. \cite{mdc98} reveal an IR and radio outburst, which is consistent
with synchrotron emission from an expanding cloud of relativistic
electrons.  The onset of the outburst is synchronized with the X-ray
spike.  At present it is unclear how the material in the accretion
disk is converted into outflowing material, or how much of the mass is
actually ejected.

%Eikenberry
%et al. (1998) obtained infrared observations simultaneous with a
%different RXTE observation, from which they draw similar conclusions. 

\section*{Variable Frequency QPO}

The power spectrum also shows a distinct change during the dip episode
(shown in the bottom panel of Figure~\ref{xte1915fig1}, with highest
powers being the darkest).  A variable-frequency QPO appears between
1100~s and 1600~s, starts at a high of $\sim$12--14~Hz, dips to
2--3~Hz, and then recovers to higher frequencies before disappearing
at the X-ray spike.  The X-ray spike seems to trigger a significant
``quiet'' period in the source variability from 1600--2000~s, after
which low frequency ($\le 4$~Hz) noise begins to dominate, as it does
during most of the rapid oscillations.  The appearance of the QPO
seems to be governed by the presence of a hard power law component
(photon index $< 3$) and low disk temperatures ($T < 1.7$~keV).  The
QPO actually reappears very briefly at about 8~Hz as a ``U''-shaped
feature near 2700~s, when these spectral conditions are satisfied.
The frequency itself is clearly correlated with intensity; preliminary
work with the dip QPO suggests that the correlation is with the disk
luminosity rather than the luminosity of the power law component.

The 67~Hz QPO \cite{mrg97} has been interpreted in terms of phenomena
occurring near the radius of marginally stable orbits.  If the lower
frequency 2--15~Hz QPOs are associated with the disk, and the
frequency scales with the distance from the black hole as the Kepler
frequency does, then they should come from a radius on the order of
$3-10$ times larger than the smallest radii observed, that is $\ge
120$~km. The spectral values for the radius of the disk when these QPO
are present are not quite this large, although of similar magnitude.
The low and high frequency QPOs may however be quite distinct.  For
example, the luminosity dependence of the frequency and the presence
of a second harmonic are properties not seen in the 67~Hz feature.
The discussion by Chen et al. \cite{cts97} applies to these QPO seen
during the transient dips as well as to the QPO seen during extended
low states. It is also possible that disk variations, oscillations, or
precession might modulate the hard X-ray flux to give large amplitude
variations of the power law component, which dominates the flux during
the dips.

% While the lower frequency would be characteristic of phenomena at
% larger radii,

\end{document}